\documentstyle[aps,twocolumn]{revtex}
\input{epsf}
\tighten

\begin{document}

\twocolumn[\hsize\textwidth\columnwidth\hsize\csname @twocolumnfalse\endcsname
\title{Standard Quantum Limits for broadband position measurement}
\author{Hideo Mabuchi}
\address{Norman Bridge Laboratory of Physics 12-33, California Institute of Technology,
Pasadena CA 91125}
\date{January 28, 1998}
\maketitle
\begin{abstract}
I utilize the Caves-Milburn model for continuous position measurements
to formulate a broadband version of the Standard Quantum Limit (SQL) for
monitoring the position of a free mass, and illustrate the use of Kalman
filtering to recover the SQL for estimating a weak classical force that
acts on a quantum-mechanical test particle under continuous observation.
These derivations are intended to clarify the interpretation of SQL's in
the context of continuous quantum measurement.
\end{abstract}
\vspace{3ex}
]

Substantial efforts have been devoted to elaborating Standard Quantum Limits
(SQL's) for both discrete \cite{Yuen83,Cave85,Ozaw88,Brag92} and continuous 
\cite{Brag92,Jaek90} measurement of the position of a free mass. In large
part, the motivation for such investigations stems from a pressing need to
identify any possible constraints imposed by the principles of quantum
measurement on the experimental possibility to detect gravitational waves 
\cite{Brag74,Cave80}. My objectives in this paper will be to formulate an SQL
appropriate to broadband continuous measurements of the position of a free mass
(expressed in terms of the signal bandwidth and with measurement sensitivity
expressed in units of length per $\sqrt{{\rm Hz}}$), to consider its proper
intepretation, and to demonstrate its compatibility with the usual SQL
\cite{Brag92} for detecting a weak classical force. The mathematical analysis
will be adapted to the continuous measurement model of Caves and Milburn \cite
{Cave87}, which appears to have direct relevance for concrete experimental
scenarios such as atomic force microscopy \cite{Milb94} and cavity QED \cite
{Dohe98,Mabu98}.

Although the limits I\ derive will be familiar from previous studies of the
discrete and narrowband measurement scenarios, my emphasis here will be on
formulating these limits in a manner that is specific to broadband quantum
measurement. In particular, I have found that some care needs to be taken in
deriving an SQL for detecting weak forces from the SQL for continuous position
measurement. The method I\ use below emulates the classical technique of Kalman
filtering \cite{Jaco93,Milb96}, and was motivated by the general strategy of
quantum system identification discussed in \cite{Mabu96c}. Throughout this paper
I have adopted the time-domain, state-space perspective of quantum measurement
whose virtues have become apparent from recent advances in adaptive quantum
measurement \cite{Wise95} and quantum feedback \cite{Wise94}. Such methods hold
great promise for the challenge of formulating experimentally tenable strategies
for broadband quantum nondemolition measurement.

The mathematical formalism necessary to treat continuous quantum
measurements has been developed by numerous authors, with the most relevant
works for the present discussion being \cite{Cave87,Milb96,Bela92}. While it
is not absolutely essential to go to the continuous limit, doing so will
allow us to use the convenient notation of stochastic differential equations
(SDE's) \cite{Gard90}. What really matters for the discussion at hand is an
assumption that the timescale associated with measurement interactions and
readouts is much shorter than any timescale on which we wish to undertstand
the system dynamics.

In the Caves-Milburn model for continuous quantum-mechanical measurements of
position \cite{Cave87} there is one system of interest, and an infinite
succession of identical (and identically-prepared) ``meters.'' Let the
meters be labelled by an index $r.$ The system is brought into momentary
contact with the $r^{th}$ meter at time $t_{r}=r\tau $, and the position
operator of each meter is measured sharply just after it has interacted with
the system. The string of measurement results thus generated constitutes a
classical record of the system evolution. Continuous measurement is achieved
in the limit where the time-interval between measurements $\tau $ goes to
zero, with the uncertainty of each individual measurement simultaneously
going to infinity in an appropriate manner.

Let $x,p$ denote the system position and momentum operators, and $%
X_{r},P_{r} $ the position and momentum operators of the $r^{th}$ meter. We
adopt the following time-dependent interaction Hamiltonian between the
system and the meters: 
\begin{equation}
H_{int}=\sum_{r=1}^{n}\delta \left( t-r\tau \right) xP_{r}.  \label{eq:Hint}
\end{equation}
The delta-function form is chosen to facilitate the limit $\tau \rightarrow
0,$ but as discussed above it may be viewed as the idealization of any
``shrinkable'' function with compact support. We assume that each meter,
just before it interacts with the system, is prepared in the pure state $%
\left| \Upsilon _{r}\right\rangle $ with Gaussian wave-function $\left(
X_{r}\left| \xi _{r}\right\rangle \equiv \xi _{r}\left| \xi
_{r}\right\rangle \right) $%
\begin{equation}
\left\langle \,\xi _{r}\,|\,\Upsilon _{r}\,\right\rangle \equiv \Upsilon
\left( \xi _{r}\right) =\frac{1}{\left( \pi \sigma \right) ^{1/4}}\exp
\left[ \frac{-\xi _{r}^{2}}{2\sigma }\right] .  \label{eq:Upsilon}
\end{equation}
Caves and Milburn have have derived an exact expression for the conditional
evolution of the system state under the following measurement protocol: at
each time $t_r$, $i.$ couple the system to the $r^{th}$ meter prepared in
state (\ref{eq:Upsilon}), $ii.$ evolve the system and meter under the
interaction Hamiltonian (\ref{eq:Hint}), $iii.$ perform a precise measurement
of $X_{r}$. If we write $\rho \left( t_{r}-\right)$ for the system state just
before the measurement at time $t_{r}\equiv r\tau$, the post-measurement
system state $\rho\left(t_r+\right)$ is given by
\begin{eqnarray}
\rho \left( t_{r}+\right) &=&\hat{\Upsilon}\left( \xi _{r}\right) \rho
\left( t_{r}-\right) \hat{\Upsilon}^{\dag }\left( \xi _{r}\right),
\label{eq:Qopform} \\
\hat{\Upsilon}\left( \xi _{r}\right) &\equiv &\left\langle \,\xi
_{r}\,\right| \,e^{-ixP_{r}/\hbar }\,\left| \,\Upsilon _{r}\,\right\rangle,
\label{eq:Qopactual}
\end{eqnarray}
where $\xi_r$ corresponds to the actual outcome of the meter projection in
stage $iii$. Recall that $x,X_{r,}P_{r}$ are Hermitian operators, $\xi _{r}$
is a c-number, and $\left| \,\Upsilon _{r}\,\right\rangle $ is a Hilbert-space
ket. $\hat{\Upsilon}_{r}$ is thus a quantum operation density.  Qualitatively
speaking, the basic effect of (\ref{eq:Qopform}) is to {\em shift} the centroid
$x_{r}^{\prime }\equiv {\rm Tr}\left[ x\rho \left( t_{r}-\right) \right] $ of
the system's position-space distribution {\em towards} the value conveyed by 
$\xi _{r},$ and to reduce its overall width $\Delta _{r}^{\prime }\equiv 
{\rm Tr}\left[ x^{2}\rho \left( t_{r}-\right) \right] -\left( x_{r}^{\prime
}\right) ^{2}$ by an amount that depends on the ratio $\Delta _{r}^{\prime
}/\sigma .$ Note that this is not a projective measurement---the
post-measurement state $\rho \left( t_{r}+\right) $ depends on both the
measurement result $\xi _{r}$ {\em and} the pre-measurement state $\rho
\left( t_{r}-\right) .$

The operation (\ref{eq:Qopactual}) maps Gaussian pure states of the system
to Gaussian pure states, so we can in fact parametrize the selective
evolution of a free particle (initially prepared in a Gaussian pure state)
by just four real numbers. Chosing $x_{r}\equiv \left\langle x\right\rangle
,p_{r}\equiv \left\langle p\right\rangle ,\Delta _{r}\equiv 2\,\left\langle
\left( \Delta x\right) ^{2}\right\rangle ,$ and $\varepsilon _{r}\equiv
\left\langle \Delta x\Delta p-\Delta p\Delta x\right\rangle /\hbar ,$ Caves
and Milburn derived a set of difference equations for the evolution of these
quantities with each successive measurement. They further showed that
Gaussian initial states generally converge toward stationary evolutions, in
which $x_{r}$ and $p_{r}$ evolve stochastically but $\Delta _{r}$ and $%
\varepsilon _{r}$ vary periodically---during the time intervals between
measurements both widths increase according to the free Hamiltonian, but the
effect of each measurement is to reduce them by a constant ``contraction
factor'' $C$ such that $\Delta _{r}\equiv {\rm Tr}\left[ x^{2}\rho \left(
t_{r}+\right) \right] -\left( x_{r}\right) ^{2}$ is independent of $r.$ The
form of the evolution equations is especially convenient in this stationary
regime, so in what follows we shall assume that stationarity has been
bootstrapped by preparing the system in an appropriate initial state.

\strut In order to derive a broadband SQL for position measurements, we
start with the stationary difference equations (3.30a,b) from \cite{Cave87}: 
\begin{eqnarray}
x_{r+1}-x_{r} &=&p_{r}\frac{\tau }{m}+\frac{C-1}{C}\left( \xi
_{r}-\,x_{r}^{\prime }\right) ,  \nonumber \\[0.05in]
p_{r+1}-p_{r} &=&\frac{\hbar }{\sigma \sqrt{C}}\left( \xi
_{r}-\,x_{r}^{\prime }\right) .  \label{eq:difference}
\end{eqnarray}
Here unprimed quantities $x_{r},p_{r}$ describe the system state just after
the $\left( r-1\right) ^{th}$ measurement has been made, and $x_{r}^{\prime
} $ holds just before the $r^{th}$ measurement.

We first need to rewrite the difference equations (\ref{eq:difference}) as a
stochastic differential equation (SDE), by taking $\tau \rightarrow 0$
together with $\sigma \rightarrow \infty $ such that their product $D\equiv
\sigma \tau $ stays constant. Recall, from equations (3.20,3.21) of \cite
{Cave87}, that $\left( \xi _{r}-\,x_{r}^{\prime }\right) $ is a Gaussian
random variable with zero mean and variance $\sigma C/2.$ In the continuous
limit it also turns out that $C\rightarrow 1$ \cite{Cave87}, so we
immediately have an explicit (Ito) stochastic differential equation: 
\begin{equation}
d\left(\matrix{x \cr p}\right) = \left(\matrix{p/m \cr 0}\right) \,dt+ \left(%
\matrix{0\,\cr \sqrt{\hbar^2\over 2D}}\right) \,dW,  \label{eq:SDE}
\end{equation}
where $dW$ is a Wiener increment \cite{Gard90}.

To derive an SDE for the rms position, we begin by applying Ito's formula to
derive an SDE for $x_{ms}\equiv \left\langle x^{2}\right\rangle ,$ 
\begin{eqnarray}
d\left( x^{2}\right) &=&\frac{2p}{m}\,x\,dt,  \nonumber \\
\frac{d}{dt}\,x_{ms} &=&\frac{2}{m}\,\left\langle \,px\right\rangle .
\end{eqnarray}
Substituting $x,p$ by integrals of equation (\ref{eq:SDE}), and keeping in
mind that the ``functions'' $dW\left( t\right) $ appearing in the
expressions for $x$ and $p$ will be identical for any given stochastic
realization, 
\begin{eqnarray}
\frac{d}{dt}\,x_{ms} &=&\frac{\hbar ^{2}}{m^{2}D}\,\left\langle
\,\int_{0}^{t}dW\left( t^{\prime }\right) \int_{0}^{t}dt^{\prime \prime
}\int_{0}^{t^{\prime \prime }}dW\left( t^{\prime \prime \prime }\right)
\right\rangle  \nonumber \\
&=&\frac{\hbar ^{2}}{2m^{2}D}\,t^{2}.
\end{eqnarray}
Hence, 
\begin{equation}
x_{rms}\left( t\right) =\frac{\hbar }{m\sqrt{6D}}\,t^{3/2}.
\end{equation}

We now turn to quantify the fundamental measurement noise inherent in the
present scheme. As mentioned above, the discrete measurement errors (before
taking the continuous limit) are a stationary Gaussian process with variance 
$C\sigma /2.$ Hence, in any given time interval $B^{-1}$ the sample variance
for the $N\equiv \left( B\tau \right) ^{-1}$ measurements will be 
\begin{eqnarray}
\left\langle \sum \xi _{r}^{2}\right\rangle &=&\frac{C\sigma }{2B\tau }, 
\nonumber \\
\Xi _{rms} &\equiv &\frac{1}{N}\xi _{rms}=\sqrt{\frac{C\sigma B\tau }{2}}%
\rightarrow \sqrt{\frac{DB}{2}}.
\end{eqnarray}
Note that $\sqrt{D}$ thus represents a measurement inaccuracy (or
``sensitivity,'' depending on your point of view) in units such as ${\rm %
meters}/\sqrt{{\rm Hz}}.$ In the sequel we shall regard $\Xi \left( t\right) 
$ as the time-domain measurement signal, so that $B$ should be interpreted
as a low-pass {\em bandwidth}. It is important to understand that $D$ and $B$
are completely independent variables---$D$ parametrizes the system-meter
coupling strength (a physical quantity), $B$ the degree of smoothing applied
to the measurement results (a signal-processing quantity).

We can now formulate a condition for backaction-induced diffusion to become
visible against the fundamental measurement noise: 
\begin{eqnarray}
x_{rms} &=&\Xi _{rms}\Rightarrow  \nonumber \\
\frac{\hbar }{m\sqrt{6D}}\,t_{*}^{3/2} &=&\sqrt{DB/2},  \nonumber \\
t_{*} &=&\left[ \sqrt{3B}\frac{Dm}{\hbar }\right] ^{2/3}.  \label{eq:bbsql}
\end{eqnarray}
This expression should be taken to predict the overall measurement time $%
t_{*}$ at which backaction-diffusion
\begin{figure}[tb]
\centerline{\epsfbox{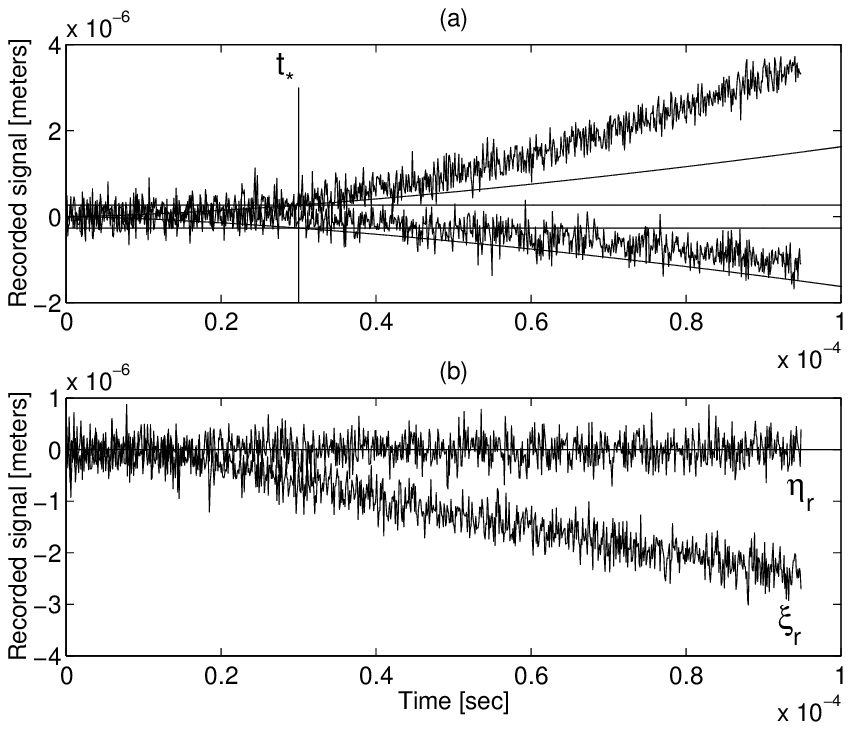}}
Figure 1. (a) Horizontal lines indicate the rms signal noise $\pm\sqrt(DB/2)$,
curved lines indicate $\pm x_{rms}(t)$, and the vertical line indicates $t_*$
for $D=1.42\times 10^{-20}$, $B=10^7$, and $m=2.22\times 10^{-25}$ (mks units).
Two stochastic realizations of the measurement signal are also shown, generated
by direct simulation of equations (19)-(21). (b) The nonstationary component of
the signal $\xi_r$ can be perfectly subtracted away to yield $eta_r\equiv\xi_r
-{\hat x}'_r$, which is a Gaussian noise process with variance $\sigma C/2$.
\end{figure}
should cause the rms wander $x_{rms}$
of the observed particle (with mass $m$) to equal the point-to-point scatter 
$\Xi _{rms}$ in continuous measurements made with measurement inaccuracy $%
\sqrt{D}$ and filtered to bandwidth $B$. Viewed as such, I\ claim that
equation (\ref{eq:bbsql}) represents the most sensible form for a {\em %
broadband} SQL\ for continuous observation of the position of a free mass. 
Note that the timescale for backaction noise to manifest itself increases
with increasing signal bandwidth (measurement noise hides the backaction),
increases with particle mass (scenario is less ``quantum''), and increases
with measurement inaccuracy (need a good measurement to get to the SQL).
This picture is illustrated by Figure 1a (for fixed $D,$ $B,$ and $m$),
which shows $x_{rms}\left( t\right) ,$ $\Xi _{rms},$ and $t_{*}$ together
with two stochastic realizations of $x\left( t\right) $.

In the special case where we choose to integrate the measurement results
over all of a given observation interval $t,$ $B\equiv 1/t$ and we recover a
single measurement scenario. We can then derive a basic figure-of-merit for
continuous measurements---defining $S\equiv \sqrt{D}$ as the measurement
inaccuracy and keeping $t$ as the total observation time, we find that we
need 
\begin{equation}
\frac{S}{t}\le 3^{-1/4}\sqrt{\frac{\hbar }{m}}  \label{eq:Sovert}
\end{equation}
in order for noise in the measurement record ultimately to be dominated by
real displacements of the particle due to backaction diffusion. Alternatively,
we can write down something like an inference-disturbance relation: 
\begin{equation}
\Xi _{rms}\cdot x_{rms}\ge \sqrt{\frac{D}{2t}}\cdot \frac{\hbar }{m\sqrt{6D}}
\,t^{3/2}=\frac{\hbar t}{2\sqrt{3}m}.
\end{equation}
This relation asserts that the product of the integrated measurement
accuracy and the rms displacement of the test particle due to
backaction-diffusion must exceed a lower bound that increases linearly with
time.

We now turn to consider the estimation of weak forces acting on the test
particle, based on something like Kalman filtering of the measurement
record. We begin by writing down a modified form of the stationary
difference equations (still explicit), 
\begin{eqnarray}
x_{r+1}-x_{r} &=&p_{r}\frac{\tau }{m}+\frac{C-1}{C}\left( \xi
_{r}-\,x_{r}^{\prime }\right) , \\[0.05in]
p_{r+1}-p_{r} &=&\alpha \tau +\frac{\hbar }{\sigma \sqrt{C}}\left( \xi
_{r}-\,x_{r}^{\prime }\right) .
\end{eqnarray}
Here we have simply added a uniform force field $\alpha $ that acts on the
particle at all times\cite{fn1}. Our task will be to detect $\alpha \ne 0$.

Based on the measurement results $\xi _{r},$ we want to update recursive
estimators $\hat{x}_{r}$ and $\hat{p}_{r}$ for $x_{r}$ and $p_{r},$
respectively. Assuming noiseless readout of $\xi _{r}$, the update equations
read 
\begin{eqnarray}
\hat{x}_{r+1}-\hat{x}_{r} &=&\hat{p}_{r}\frac{\tau }{m}+\frac{C-1}{C}\left(
\xi _{r}-\,\hat{x}_{r}^{\prime }\right) , \\[0.05in]
\hat{p}_{r+1}-\hat{p}_{r} &=&\frac{\hbar }{\sigma \sqrt{C}}\left( \xi _{r}-\,%
\hat{x}_{r}^{\prime }\right) .
\end{eqnarray}
Note that we do not include $\alpha $ in the update rules, as it is assumed
to be an unknown quantity. In order to infer $\alpha $ from the recursive
estimators $\hat{x}_{r}$ and $\hat{p}_{r}$, we must focus on the behavior of
the quantity $\eta _{r}\equiv \xi _{r}-\,\hat{x}_{r}^{\prime }.$ As we know
the distribution for $\xi _{r},$ we can write 
\begin{equation}
\eta _{r}=\left( x_{r}^{\prime }-\hat{x}_{r}^{\prime }\right) +\sqrt{\frac{%
C\sigma }{2}}{\cal N}_{r}\left[ 0,1\right] ,
\end{equation}
where ${\cal N}_{r}\left[ 0,1\right] $ is a Gaussian deviate with zero mean
and unit variance. Hence, $\eta _{r}$ will generally be the sum of a
Gaussian deviate and an uncorrelated process (the ``error-signal'') that
reflects the cumulative inaccuracy of our recursive position estimator.

Note that if there is no external force acting on the particle ($\alpha =0$)
there is no measurement inaccuracy, and the recursive estimator $\hat{x}_{r}$
can be used {\em perfectly} to subtract the backaction-induced diffusion from
the signal $\xi _{r}$ (see Figure 1b). Indeed, demonstrating the ability to do
so would seem to be the only good way of verifying that an actual experimental
broadband position measurement reaches the ``quantum regime'' defined by
equation (\ref{eq:Sovert})---unlike the scenario of optical measurements of
quadrature amplitudes, there is no way directly to compare signal and meter
beams to determine the fidelity of the measurement. If there is external force,
the subtraction will not be perfect and $\alpha$ will reveal itself in
nonstationary behavior of $\eta _{r}.$

Note as well that the variance of the residual noise process $\sqrt{\frac{C\sigma}
{2}}{\cal N}_r\left[ 0,1\right] $ can in principle go to {\em zero} in the limit of
good measurements. At first thought this might seem to have profound implications
for the detection of weak classical forces, but in fact the quantitative analysis
below recovers the usual SQL for weak force detection.

Before proceeding any farther, let us make the transition to the continuous limit
($\tau \rightarrow 0,$ $\sigma \rightarrow \infty ,$ $D={\rm const}$) \cite{Milb96}
and collect together a complete set of SDE's for the physical quantities and
statistical estimators: 
\begin{eqnarray}
dx &=&\frac{p}{m}\,dt,\qquad dp=\alpha dt+\sqrt{\frac{\hbar ^{2}}{2D}}dW, \\
d\xi &=&xdt+\sqrt{\frac{D}{2}}dW, \\
d\hat{x} &=&\frac{\hat{p}}{m}dt,\qquad d\hat{p}=\frac{\hbar }{D}\left( d\xi -%
\hat{x}dt\right) .  \label{eq:modelend}
\end{eqnarray}
In these terms, $d\xi /dt$ represents the broadband measurment record. Note
that we could also define an auxilliary variable $e\equiv x-\hat{x},$ in
terms of which 
\begin{eqnarray}
de &=&\frac{1}{m}\left( p-\hat{p}\right) dt, \\
d\hat{p} &=&\frac{\hbar }{D}\,e\,dt+\sqrt{\frac{\hbar ^{2}}{2D}}dW.
\end{eqnarray}
Our estimator for the external force $\alpha $ will be proportional to the
time integral of $d\eta \equiv d\xi -\hat{x}dt,$ so what we really need to
know is the rms behavior of $e\left( t\right) $. Developing its SDE, 
\begin{eqnarray}
de &=&dx-d\hat{x},  \nonumber \\
\frac{de}{dt} &=&\frac{1}{m}\left( \alpha t-\frac{\hbar }{D}%
\int_{0}^{t}e\left( t^{\prime }\right) \,dt^{\prime }\right) ,  \nonumber \\
\frac{d^{2}e}{dt^{2}} &=&\frac{1}{m}\left( \alpha -\frac{\hbar }{D}%
\,e\right) .
\end{eqnarray}
This shows that $e\left( t\right) $ behaves like a harmonic oscillator, with
mass $m$ and natural frequency $\omega _{0}=\sqrt{\frac{\hbar }{mD}}$,
subjected to a constant external force $\alpha .$ So with the initial
condtion $\hat{x}=x,$ $\hat{p}=p,$ we expect 
\begin{equation}
e\left( t\right) =\frac{\alpha D}{\hbar }\left[ 1-\cos \left( \sqrt{\frac{%
\hbar }{mD}}t\right) \right] .
\end{equation}
Getting back to the time integral of $d\eta ,$%
\begin{equation}
\int_{0}^{t}\left( \frac{d\xi }{dt^{\prime }}-\hat{x}\right) dt^{\prime }
=\int_{0}^{t}e\left( t^{\prime }\right) dt^{\prime }+\sqrt{\frac{D}{2}}%
\int_{0}^{t}dW\left( t^{\prime }\right) \\
\end{equation}
from which we identify the signal $\Sigma $ and rms noise $N$ as 
\begin{equation}
\Sigma =\frac{\alpha D}{\hbar }\left[ t-\sqrt{\frac{mD}{\hbar }}\sin \left( 
\sqrt{\frac{\hbar }{mD}}t\right) \right] ,\quad N=\sqrt{\frac{Dt}{2}}.
\end{equation}
Setting $\Sigma /N=1$ we obtain 
\begin{equation}
\alpha _{\min }=\hbar \sqrt{\frac{t}{2D}}\left[ t-\sqrt{\frac{mD}{\hbar }}%
\sin \left( \sqrt{\frac{\hbar }{mD}}t\right) \right] ^{-1}.\label{eq:amin}
\end{equation}
\newline
For fixed $m$ and $t$, it seems reasonable that the optimal choice of $D$
should be simply related to $\hbar t^{2}/m.$  If we substitute $D\equiv
\eta^2\hbar t^2/m$ into (\ref{eq:amin}) and then (numerically) minimize over
$\eta$, 
\begin{eqnarray}
\alpha _{\min } &=&\min_{\eta }\sqrt{\frac{\hbar m}{2t^{3}}}\left[ \eta
-\eta ^{2}\sin \left( 1/\eta \right) \right] ^{-1}\approx\pi\sqrt{\frac{
\hbar m}{2t^{3}}}.
\end{eqnarray}
Hence we are able to recover (within a factor of $\pi $) the usual SQL for
detecting a weak classical force via repeated position measurements on a
free mass \cite{Brag92}.

The author\ would like to thank J. W. Hartman and Y. Levin for discussions,
and to acknowledge support from the National Science Foundation under Grants
No. PHY94-07194 and \strut PHY97-22674.


\begin{references}
\bibitem{fn1}  Equations (14)-(15) can be derived by direct solution of the
Schr{\"{o}}dinger equation with a Gaussian ansatz.
\end{references}
\end{document}